\documentclass[12pt]{article}
\usepackage[superscript]{cite} 
\usepackage{times} 
\usepackage{hyperref}
  \usepackage{multirow}
\usepackage{geometry}
\usepackage{color}
\usepackage{array}

\usepackage[T1]{fontenc}  \usepackage{graphicx}  
\usepackage{breakcites} 
\usepackage[margin=0.5cm]{caption}

\newcommand{\itemb}{\begin{itemize}}
\newcommand{\iteme}{\end{itemize}}
\newcommand{\msection}[1]{\normalsize \section{{#1}}}  \newcommand{\msubsection}[1]{\normalsize \subsection{{#1}}}  \newcommand{\smallest}[1]{\vspace{6pt}\noindent {\bf {#1}.}}  
 \newcommand{\me}{\bf}

\newcommand{\indentparb}{\begin{adjustwidth}{2cm}{}}
\newcommand{\indentpare}{\end{adjustwidth}}

\newcommand{\TAD}{{{T*AD}}}  \newcommand{\ptname}{{{T*PL}}}  
\newcommand{\cgen}{{{Cgen}}}
\newcommand{\cres}{{{Cres}}}

\newcommand{\degen}{{{degeneration}}}
\newcommand{\destab}{{{destabilization}}}
\newcommand{\potn}{{{potentiation}}}
\newcommand{\stab}{{{stabilization}}}
\newcommand{\consol}{{{consolidation}}}
\newcommand{\Consol}{{{Consolidation}}}

\newcommand{\cholest}{{{cholesterol}}}
\newcommand{\Cholest}{{{Cholesterol}}}

\newcommand{\lrf}{{{lipid raft}}}

\newcommand{\cskl}{{{cytoskeleton}}}

\newcommand{\mtb}{{{microtubule}}}

\newcommand{\nanod}{{{nanodomain}}}

\newcommand{\Ecm}{{{Extracellular matrix}}}

\newcommand{\icell}{{{intracellular}}}
\newcommand{\ecell}{{{extracellular}}}

\newcommand{\metab}{{{metabolism}}}

\newcommand{\infl}{{{inflammation}}}
\newcommand{\phos}{{{phosphorylation}}}
\newcommand{\Phos}{{{Phosphorylation}}}

\newcommand{\constit}{{{constitutive}}}
\newcommand{\Constit}{{{Constitutive}}}

\newcommand{\astc}{astrocyte}

\newcommand{\Astc}{Astrocyte}

  \newcommand{\hippo}{hippocampus}

\newcommand{\plast}{{{plasticity}}}
\newcommand{\Plast}{{{Plasticity}}}
\newcommand{\transm}{{{transmission}}}

\newcommand{\mitoch}{{{mitochondria}}}
\newcommand{\Mitoch}{{{Mitochondria}}}

\newcommand{\Ngns}{{{Neurogenesis}}}

\newcommand{\catwo}{{{Ca}\textsuperscript{{\raisebox{-1pt}{$\scriptstyle{2+}$}}}}}

\usepackage{lineno}  \usepackage{tikz}  \usetikzlibrary{arrows}  \usepackage{changepage}

\begin{document}
\title{A Lipid Rafts Theory of Alzheimer's Disease}

\author{
{Ari Rappoport}\\ The Hebrew University of Jerusalem\\{ari.rappoport@mail.huji.ac.il}
 }
\date{November 2020\\ August 2023, May 2024\\ \vspace{0.5cm}} 
\maketitle
\begin{adjustwidth}{30pt}{30pt}
{\bf Abstract.} 
 I present a theory of Alzheimer's Disease (AD) that explains its symptoms, pathology, and risk factors. To do this, I introduce a new theory of brain plasticity that elucidates the physiological roles of AD-related agents. 
New events generate synaptic and branching candidates competing for long-term enhancement. Competition resolution crucially depends on the formation of membrane \lrf s, which requires \astc-produced \cholest. 
Sporadic AD is caused by impaired formation of plasma membrane \lrf s, preventing the conversion of short- to long-term memory, and yielding excessive tau \phos, \icell\ \cholest\ accumulation, synaptic dysfunction, and neuro\degen. 
Amyloid beta (Abeta) production is promoted by \cholest\ during the switch to competition resolution, and \cholest\ accumulation stimulates chronic Abeta production, secretion, and aggregation. 
The theory addresses all of the major established facts known about the disease, and is supported by strong evidence.  

\smallest{Keywords} Alzheimer's disease (AD), adaptive response \plast\ (\ptname ), double-edged \plast\ (DEP), short-term memory, \cholest, \lrf s, amyloid beta.
\end{adjustwidth}

 \msection{Introduction}
 Alzheimer's Disease (AD) is the main type of dementia, affecting dozens of millions of people \cite{AZ2023}.
    Despite extensive research (215K papers in Pubmed, 1.9M in Google Scholar) and many hypotheses \cite{liu2019his}, 
                there is at present no coherent theory of AD. The dominant account is the amyloid beta (Abeta) hypothesis, 
     but it is not a complete theory and it is widely criticized \cite{morris2018abeta}.

The minimal requirement from any theory of AD is to explain the following major facts.  The main AD symptom is memory impairment, starting with anterograde memory and progressing to retrograde memory. Other cognitive deficits are also present \cite{AZ2023}. 
  Olfaction and hearing problems occur very early, as are retinal abnormalities \cite{murphy2019olf}.
  There are two disease variants, sporadic and autosomal dominant (familial (FAD)), with the former appearing only in relatively old age. 
The defining AD pathologies are intracellular neurofibrillary tangles (NFTs) associated with hyperphosphorylated tau (hyper p-tau), and extracellular plaques containing Abeta \cite{AZ2023}. 
 Tau pathology shows good spatial and temporal correlation with symptoms \cite{kaufman2018tau}, 
 but plaques are also commonly present in non-demented people in aging, and not all demented people exhibit them \cite{morris2018abeta}.
   Several basic cellular components show dysfunctions that often precede NFTs and plaques, including in \mitoch\ and lipid (especially \cholest) and calcium homeostasis \cite{area2017alz}.
        AD involves early synaptic damage and loss, followed by severe neuro\degen\ \cite{crews2010molecular}. 
  Carrying the apolipoprotein ApoE4 isoform is strongly associated with AD, while ApoE2 is almost completely protective.
There are other risk factors (female sex, viral infection, brain injury, insulin resistance, inflammation, damaged brain blood vessels, stress),  and protective factors (education). Obesity and being underweight increase risk, but being overweight is protective. 
 
   This paper presents a new theory of AD, the {\bf lipid rafts theory of AD (\TAD)}, that explains all of these facts. To do so, it first presents a new theory of brain \plast, {\bf adaptive response \plast\ (\ptname )}, which explains how neural activity yields memory formation, and the physiological roles of the main AD agents. 
The focus on \plast\ is motivated by the fact that initial memory symptoms involve anterograde amnesia of benign everyday events \cite{AZ2023}. This points to an early impairment in normal memory formation processes, i.e., in brain \plast. 
   Current descriptions of brain \plast\ are incomplete and do not address AD agents, prompting the development of a new theory. 

\ptname\ and \TAD\ address the major established facts known about brain \plast\ and AD, methodically identified via a thorough examination of the scientific literature (hundreds of thousands of papers) done over more than ten years. Both theories are supported by very strong evidence. 

\begin{figure*}
\begin{center}
\includegraphics[width=5in]{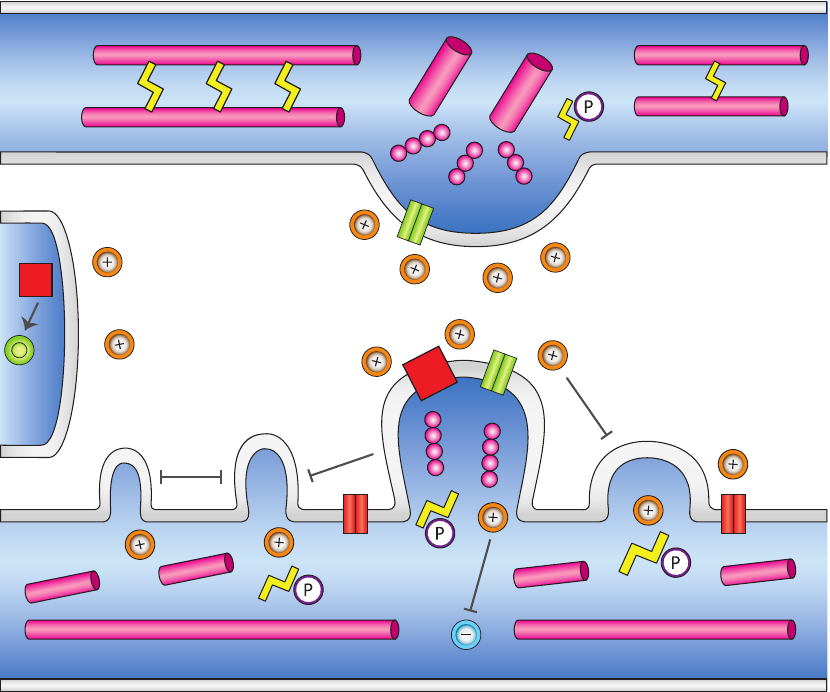}

\caption{{\bf Candidate generation (\cgen). }
Neural activity triggers the release and activation of \cgen\ agents (e.g., tPA, MMP9, BDNF; $+$ circles). These 
(i) induce tau (yellow) \phos\ (P), which destabilizes cytoskeleton microtubules (pink-purple cylinders) to allow candidate generation, 
(ii) induce the formation of competing candidates (four postsynaptic dendritic spines, one presynaptic axon bouton, calcium-permeable glutamate receptors (red cylinders)) supported by dynamic actin (pink-purple circles), 
(iii) suppress \cres\ agents ($-$ circle), 
and
(iv) promote candidate growth. sAPPa (red square) is located outside lipid rafts (light green cylinders) and promotes \cgen. In \astc s, it promotes \cholest\ synthesis (green circle).
// Upper part: presynaptic neuron. Lower part: postsynaptic neuron. Left part: \astc.
}
\label{fig1}
\end{center}
\end{figure*}

 \smallest{Overview of adaptive response \plast\ (\ptname )}
      In \ptname, the organism can be in one of three states.   The {\bf steady state} of neurons, their interconnections (synapses), and glia is supported by stable receptors and channels, the \cskl, adhesion and scaffolding molecules, trans-synaptic \nanod s, extracellular matrix (ECM), myelin, and various \constit ly active enzymes. 
Novelty-induced neural activity initiates a \plast\ process comprising two stages. First, a {\bf candidate generation (\cgen)} stage involves the formation of elements competing for long-term enhancement. Candidates include existing synapses, new synapses, new neurite branches, and (sometimes) new neurons. This stage involves \destab\ of the steady state, immediate \potn\ of neuro\transm, and structural changes, and is what underlies short-term memory (STM) (Figure \ref{fig1}). Second, {\bf competition resolution (\cres)} comprises the enhancement of winners, elimination of losers, and re\stab\ (Figure \ref{fig2}). 
 Winner enhancement starts early, but involves a relatively long process in which the evoked changes undergo \consol\ to establish a new steady state (Figure \ref{fig3}). 

Crucially, the switch from \cgen\ to \cres\ depends on the formation of plasma membrane (PM) {\bf \lrf s (LRs)}, ordered lipid nanodomains enriched in \cholest, sphingomyelin, and phospholipids, which are essential for connecting the membrane to the \cskl, the ECM, and synaptic partners, and hence for receptor trafficking and signaling, winner enhancement, and \stab.
 The role of tau is to promote \cskl\ growth and stability in winning synapses. It is activated in a PM LR-dependent manner via de\phos\ during early winner enhancement, and inactivated during \cgen\ and loser removal. 
 The role of amyloid precursor protein (APP) is to manage \cholest\ during \plast. APP has two mutually exclusive cleavage products, {\bf soluble APP alpha (sAPPa)} and Abeta. sAPPa is a major \cgen\ agent promoting \cholest\ synthesis and transport. The incorporation of sufficient amounts of \cholest\ in LRs triggers the production of Abeta, which terminates \cgen\ and removes losing candidates. 
   
 \smallest{Overview of \lrf s theory of AD (\TAD)}
Once \plast\ at health is understood, we can see what goes wrong in AD. \TAD\ provides a complete mechanistic account of AD symptoms, pathology, and risk factors. The central thesis is that AD is caused by impaired formation of plasma membrane LRs, mainly due to insufficient neural uptake of \cholest, which is normally produced by \astc s and transported to neurons via ApoE. 

Reduced formation of PM LRs prolongs \cgen\ and prevents \cres\ (Figure \ref{fig4}), with three major consequences. First, short-term memory is not converted to long-term memory, explaining AD anterograde amnesia. Second, the \cskl\ is not linked to the membrane properly, leading to excessive p-tau, \mitoch\ issues, impairment in all signaling triggered by plasma membrane receptors (including growth factor signaling), and eventual neuro\degen. 
Tau \phos\ is a direct consequence of the core mechanism of the disease, explaining the correlation between tau pathology and symptom severity. 

A third consequence of impaired plasma membrane LRs is increased \cholest\ production by neurons. In many cases, this does not suffice for PM LRs and stimulates chronic Abeta production, leading to membrane Abeta cation channels that induce calcium toxicity and to Abeta aggregation.
  In other cases, neural \cholest\ production does manage to establish PM LRs, with intact \plast\ accompanied by excessive Abeta. This explains why Abeta plaques are only weakly associated with AD symptoms and why cognitively healthy people can exhibit them. They are neither causal nor essential for the disease.

Neuro\degen\ is exacerbated by the fact that many \plast\ agents obey a {\bf double-edged \plast\ (DEP)} principle, whereby an agent can support both \cgen\ and \cres, depending on amounts or cleavages. Crucially, \catwo\ is a DEP agent, with high \catwo\ marking winners and low \catwo\ inducing loser elimination. Since the LR impairment leads to chronic \plast\ processes that involve relatively small calcium amounts, AD involves chronic loser elimination processes that yield \degen. 

Impaired formation of PM LRs can stem from several causes. In general, it occurs due to accumulation of risk factors, the strongest of which being aging. Other risk factors include ApoE4, infection by viruses that target LRs, brain insulin resistance (which impairs \cholest\ synthesis), damaged brain vasculature, brain injury, stress, and chronic neuro\infl. 

 The following sections detail \ptname, \TAD, evidence, and AD risk factors. 

 \msection{Theory of Adaptive Response \Plast\ (\ptname )}\label{sec:arp}

Brain responses with any element of novelty trigger \plast\ processes that alter the paths involved in response execution, to facilitate the activation of this response in similar future situations. \Plast\ can modify, create or remove synapses, neurite branches, neurons, glia, and glia processes. 
The fundamental \plast\ task is to identify or create the existing or new elements to be modified, perform the modification, and integrate the changes into the existing network. 
 Adaptive response \plast\ (\ptname ) explains how this is done. 

During the stable {\bf steady state}, the brain uses a variety of mechanisms to achieve execution efficiency via structural and molecular stability. When \plast\ is required, a two-stage process is initiated. The first stage, {\bf candidate generation (\cgen)}, involves de\stab\ of the steady state to allow modifications, and generation of potential elements to be modified. These include synapses to be potentiated, new neurite branches, new boutons and spines, and new neurons. Candidates are generated in all areas in which there is novelty-related execution, as indicated by high calcium influx. A single neuron normally contains many synapse candidates, to provide a variety of options for optimal \plast. 
The changes done at this stage promote high calcium influx, via membrane insertion of calcium-permeable (CP) receptors. 

Candidates compete for long-term persistence, the winners being those who have managed to gain sustained support from \icell\ calcium. Competition is race-like, not direct, although candidates do compete indirectly over limited cellular resources. 

In the second stage, {\bf competition resolution (\cres)}, winning candidates are enhanced, losing ones are eliminated, and the relevant structures are stabilized. Winner enhancement optimizes neuro\transm\ in terms of time and energy by creating spatially and temporally focused signaling, and by replacing the CP receptors added during \cgen\ by receptors that allow less or no calcium. Winning spines are structurally supported via the formation of an actin \cskl, which requires tau-mediated connectivity to \cskl\ \mtb s (MTBs), and the formation of trans-synaptic \nanod s. 
Losing spines and branches are retracted and removed, with appropriate modifications to the \cskl. There is a relatively long \consol\ process, much of it occurring during sleep,
that involves protein synthesis to re-establish membranes, \cskl s, and the ECM. 

There is overwhelming evidence that neural activity changes neuro\transm\ and induces new synapses, dendritic spines, axonal boutons, branching points, and (sometimes) neurons, that the neurons and neurites connected by new synapses participate in the inducing responses, that these changes undergo stabilization and \consol, and that this process underlies memory \cite{abraham2019plasticity}. 
                 There is also strong direct evidence for inter-spine competition \cite{stein2019spines}. 
          Further evidence is given below.

The switch from \cgen\ to \cres\ is triggered by the formation of plasma membrane LRs, which are essential for connecting the membrane to the \cskl, the ECM, and synaptic partners, and thus for trafficking and anchoring of receptors and scaffold proteins to winning candidates and for linking presynaptic and postsynaptic cells. The most sensitive element in LR formation is \cholest, because it is produced by \astc s and transported to neurons. 
Unlike other \astc-produced agents (e.g., glutamine, lactate), \cholest\ has a large hydrophobic domain and its transport requires relatively heavy machinery (ApoE). 
APP is a central \plast\ protein whose major physiological role is to manage \cholest\ during \plast, with sAPPa promoting its synthesis and Abeta terminating it. 

\ptname\ is a general theory in which \plast\ consists of a single well-defined process evoked in all \plast-inducing contexts, including neonatal development, adult learning, and after injury. It also operates in the same manner across all body tissues. These aspects will not be detailed here due to our brain and AD focus. 

 \smallest {Double-edged \plast\ (DEP)} 
The brain doesn't know in advance which candidates would win and which would lose, so both winners and losers are stimulated by the same agents. To distinguish between winners and losers, biology utilizes a mechanism that I call {\bf double-edged \plast\ (DEP)}. In DEP, different agent amounts, rates, or cleavages can induce opposite effects. To induce amount-dependent effects, the agent binds different receptors or domains with different affinities. Usually, winners and losers utilize low and high affinities (high and low amounts), respectively. 
Brain agents are chemical and propagate in media, so only a focused execution path enjoys high amounts, defining where winners are located.  
 
 Many important agents directly obey DEP. Calcium activates CaMKII/PP2B, promoting \cgen/\cres\ (see below). Dopamine has two receptor families, the low affinity D1R family that supports vigorous activity (via GPCR Gs and Gq) and the high affinity D2R family that opposes it (via Gi/o) \cite{beaulieu2015DA}. Similarly, high/low affinity serotonin 5-HT$_1$/5-HT$_2$ receptors suppress/promote effortful activity \cite{barnes1999SER}.

Another mechanism that supports DEP is different cleavages of a protein. To induce cleavage effects, the pre-cleavage and various post-cleavage agents activate different signaling paths. This is the case with APP (see below) and growth factors (BDNF, NGF), whose mature forms support candidate and winner growth while their proforms support loser removal (see below).

 \smallest{LTP, LTD}
The notions of long-term \potn\ (LTD) and long-term depression (LTD), which are ubiquitously used in relation to \plast, are not used in \ptname. LTP occurs in both \cgen\ (all over) and \cres\ (in winners), and LTD occurs in \cres\ in both winners and losers. These terms are suitable for describing the local effects of specific experimental conditions, not for the high-level theory. 
 
The following sections detail the steady state, \cgen, \cholest\ and LRs, and \cres, with an emphasis on the \ptname\ aspects that are most relevant to AD. 

 \msubsection{The steady state}
The main relevant components of the steady state are the \cskl, adhesion and scaffold proteins, the ECM, myelin, and \constit\ activity. 
 
 \smallest{Cytoskeleton}
The \cskl\ supports intracellular trafficking, including of \mitoch\ and growth factors, exocytosis and endocytosis, 
          and is crucial for neurotransmission and \plast\ \cite{bednarek2011beta}. 
  It has three main ingredients, MTBs, located mainly in axons but also in dendritic spines,  intermediate filaments, and F-actin, located in spines.    MTB and actin stability are mainly supported by Map1b and cofilin, respectively.
   Tau is discussed below. 

 \smallest{Adhesion}
A variety of synaptic adhesion proteins (e.g., neuroligin/neurexin, LRRTM) connect pre- and postsynaptic sites to improve physical stability and signaling \cite{biederer2017tra}. 
  An important role of these proteins is to establish trans-synaptic nanocolumns and \cskl-connected synaptic {\bf \nanod s} that restrict the spatial propagation of neurotransmitters to provide rapid and efficient activation of downstream signaling \cite{biederer2017tra}.
                    Stable localization of receptors at the plasma membrane is assisted by scaffold proteins such as PSD95 and GRIP1 \cite{chung2000pho}. 
    
 \smallest{\Ecm\ (ECM)}
The brain's ECM is  mainly arranged in perineuronal nets and perinodal ECM \cite{fawcett2019rol}. 
 It is synthesized following neural activity    and provides stability.     It closes the critical period of \plast\ during development, opposes the lateral diffusion of some synaptic proteins to extrasynaptic sites, 
  and augments synaptic adhesion \cite{fawcett2019rol}.    
Integrins are a family of cell surface ECM adhesion molecules that support \cskl-based bidirectional trafficking \cite{muriel2018rol}. 
  Thus, the steady state includes a global skeleton that connects all cells to each other via \cskl s, adhesion molecules, and ECM. 
       
 \smallest{Myelin}
Oligodendrocyte (OLG) processes enwrap neural axons in a myelin sheath mainly made of OLG-produced \cholest. Myelin protects axons and improves neuro\transm.   
 \smallest{\Constit\ activity}
There are many proteins showing \constit\ activity to support the steady state. The ones most relevant here are CaMKII, PKMz, and GSK3b (see below). 

 \msubsection{Candidate generation (\cgen)}
\cgen\ is triggered in acute states involving PKA activation and strong \catwo\ influx, depolarizing the cell and inducing action potentials. This is rapidly followed by the release of the growth factor proforms proBDNF and proNGF \cite{leschik2019pro} (both pre- and post-synaptic),
 tissue plasminogen activator (tPA), which cleaves proforms to the mature forms BDNF and NGF \cite{hebert2016sto}, 
    and MMP9, which digests ECM and adhesion molecules \cite{knapska2016MMP9}. 
  Cell detachment from the ECM activates integrin signaling and yields the endocytosis of LR components such as caveolin1, destabilizing the steady state \cite{muriel2018rol}. 
 
   PKA, MARK2 and high \catwo-activated CaMKII phosphorylate tau at S262, S356 and S214, and ERK phosphorylates tau at S199/S202 \cite{tapia2019tau}. 
   \Phos\ of tau inactivates it to destabilize the existing MTB-actin \cskl\ and is essential for neurite formation. 
  ERK also destabilizes myelin \cite{cervellini2018sus}. 
  Thus, \cgen\ destabilizes both extracellular and intracellular \stab\ mechanisms.

A central \ptname\ tenet is that \cgen\ is a \cholest\ deficiency state. In support, statins, which decrease \cholest\ synthesis, increase tPA and p11 (another agent that facilitates growth factor maturation) expression and BDNF maturation activity \cite{ludka2017ato}, 
 and \cholest\ deficiency induces hyper p-tau, axonal \degen, and MTB depolymerization (breakdown) \cite{fan2001cholest}. 
 NGF is a \plast\ agent supporting winners. It cooperates with the Akt axis (see below) to promote \cholest\ synthesis and rapid axonal growth \cite{zhou2004ngf}.

Silent synapses contain NMDARs (mainly GluN2B) but not AMPARs \cite{itami2003bra}. 
  BDNF acts via TrkB and enhances GluN2B currents to promote neurite branching \cite{park2013BDNF}
   and AMPARs (mainly the GluA1 subunit) trafficking to silent synapses, unsilencing them \cite{itami2003bra}. 
These unsilenced synapses are branching candidates. 
    CaMKII binding to GluN2B is needed for spinogenesis \cite{nicoll2023syn}. 
   Wnt protein, which promotes growth during development, also promotes axonal branching during \cgen\ \cite{armijo2017takes}. 
 
GluA2-lacking AMPARs are calcium permeable. CP-AMPARs are located at extrasynaptic or synaptic sites, and mark synaptic candidates. 
During \cgen, PKA and CaMKII stimulate GluA1 surface expression via \phos\ at S845 and S831 \cite{henley2013ampa}, with 
 ERK promoting GluA1 trafficking to synapses \cite{zhu2002ras}.
    Existing synapses are not degraded, but are destabilized via the digestion of adhesion molecules, and their excitatory currents are enhanced by switching AMPARs to be CP via GluA2 endocytosis promoted by PKC \phos\ at S880 \cite{bissen2019amp}. 
 
Most PKC isoforms are activated by calcium. In addition to GluA2 endocytosis, PKC promotes membrane incorporation of GluA1 via pS181 \cite{boehm2006synaptic}, 
 and activation of the Rho GTPase Rac1, which promotes spinogenesis via the actin \cskl\ \cite{pilpel2004activation}.

Reelin is an agent promoting neural migration and growth during development, after which its expression diminishes. In the adult brain, it is expressed by excitatory neurons in entorhinal cortex (ETRC) layer 2 projecting to the \hippo, GABAergic neurons in cortical layers 1/2, and retrosplenial cortex-projecting GABAergic \hippo\ neurons \cite{yamawaki2019long}. All of these neurons have axons in long-distance pathways. Reelin is a \cgen\ agent promoting APP alpha cleavage \cite{balmaceda2014apoer2} and neurite branching \cite{stranahan2013reelin}. In \ptname, its role is to promote integration of new neurons and events into the network.

 \smallest{APP, sAPPa}
Amyloid precursor protein (APP) is a transmembrane protein cleaved by alpha-secretase to yield soluble APPa (sAPPa), with the remaining part later cleaved by gamma-secretase, or by beta-secretase (BACE1) followed by gamma-secretase to generate Abeta. 
Despite extensive research on Abeta and APP \cite{vanDerKant2015cellular}, 
   to my knowledge there is no coherent account of their physiological roles. \ptname\ provides such an account, where sAPPa and Abeta are \cgen\ and \cres\ agents, respectively. A specific role of APP is to manage \cholest, with sAPPa promoting its synthesis and transport, and Abeta terminating its synthesis and promoting extrusion. 

APP transcription is mainly promoted by \cgen\ agents, including BDNF \cite{ge2002reg} 
           and ERK \cite{huang2017APP}. 
       Alpha cleavage (i.e., sAPPa production) is promoted by \cgen\ agents, including BDNF \cite{nigam2017APPalpha}, 
 NGF \cite{yang2014APPalpha}, 
 tPA \cite{ledesma2000plasmin}, 
 MMP9\cite{fragkouli2011MMP9}, 
  and PKC \cite{buxbaum1990APPalpha}. 
     Supporting the \ptname\ thesis that \cgen\ is a state in which membrane \cholest\ is needed, alpha cleavage is stimulated by reduced membrane \cholest\ \cite{kojro2001cholest}, and occurs in non-LR domains \cite{ehehalt2003APP}. 
    sAPPa promotes \cholest\ synthesis. The \icell\ part created by alpha cleavage increases SREBP2, the main agent inducing \cholest\ synthesis, in cells with low \cholest, including human \astc s \cite{wang2014APP}. 
 sAPPa directly binds to and activates insulin receptors \cite{jimenez2011APPalpha},    which activate SREBP2 via the PI3K/Akt/mTOR axis and increase alpha-secretase expression \cite{wang2014ins}.

sAPPa is a major \cgen\ agent. It rapidly increases surface extrasynaptic GluA1, decreases GluA2/GluA3 and de novo GluA2 synthesis \cite{livingstone2021sec}, 
 and promotes neurite growth, branching, spine density, and spine volume \cite{hick2015acute}. 
   It is an adhesion molecule that promotes axonal outgrowth and synaptogenesis \cite{sosa2017APP}.

\cgen\ directly opposes \cres\ by suppressing BACE1 via several \cgen\ agents including sAPPa \cite{obregon2012BACE1}, 
  NGF, TrkA \cite{fragkouli2011MMP9}, 
 and BDNF (via SORLA) \cite{rohe2009BDNF}. 
        \cgen\ does not end until winners can stably grow, which requires plasma membrane \lrf s. 

 \smallest{Short-term memory (STM)}
\cgen\ is what supports STM. In particular, it has been repeatedly shown that GluA1 is essential for STM, with knockout animals showing normal spatial reference memory but markedly impaired spatial working memory and STM \cite{gugustea2021gen}. 
  \msubsection{\Cholest, ApoE, \lrf s}
 \cgen\ is a \cholest -deficient state, and the incorporation of \cholest\ in LRs switches \plast\ to \cres. 
\Cholest\ is a vital element in all body tissues \cite{martin2010cholest}. 
   The brain is the most \cholest-rich organ, containing over 20\% of body \cholest. \Cholest\ is a major ingredient of LRs and myelin. It is also the precursor of steroids, which are major \plast\ agents. 
    \Cholest\ does not cross the blood-brain barrier. Brain \cholest\ is produced in \astc s, transported to neurons in {\bf apolipoprotein E (ApoE)} particles, and uptaken via ApoE receptors (mainly LDL and LRP1). 
ApoE transcription is promoted in growth situations via growth factors (BDNF, NGF) and estrogen 
 \cite{spagnuolo2018BDNF, strachan2015NGF, wang2006ApoE}. 
    It is normally produced by neurons only in small amounts, with production increasing in stress, injury and deficiency conditions \cite{mahley2012injury}. 
     
\Cholest\ synthesis is mainly stimulated by SREBPs \cite{shimano2017SREBP}, which also promote LDLR. 
 SREBPs are activated by lack of plasma membrane \cholest, and by ERK and insulin/Akt (see above). 
    The brain-specific \cholest\ metabolite 24SHC (and its non-brain oxysterol counterpart 27HC), and intracellular \cholest\ excess, suppress SREBP and activate nuclear LXRs, which promote ApoE and the transporters ABCA1 and ABCG1.  These extrude \cholest\ by lipidating ApoE and transporting it across the membrane.

 \smallest{ApoE isoforms}
 Humans have three ApoE isoforms that differ in produced amounts and transport capacity, with a ranking of ApoE2 > ApoE3 > ApoE4 \cite{yamazaki2019ApoE}. 
  ApoE4 has the highest receptor affinity, saturating receptors faster. These properties can have both positive and negative consequences. ApoE4 is more efficient in terms of speed and resources, explaining its various positive effects (e.g., improving cognition at youth \cite{mondadori2007ApoE4}). 
      However, the fact that neurons of ApoE4 carriers receive less \cholest\ makes them vulnerable in aging, as discussed here.

 \smallest{Lipid rafts}
 LRs are dynamic membrane nanodomains enriched in \cholest, 
sphingolipids, mainly sphingomyelin (SPM), and glycerophospholipids such as gangliosides \cite{sezgin2017mys}. 
 LRs were controversial until some time ago due to technical difficulties, but are now fully consensual. 
LRs increase local membrane rigidity, and serve for anchoring proteins and for linking the membrane to the \cskl\ and the ECM, thereby supporting the link between the \icell\ and \ecell\ environments, in particular \ecell -induced signaling \cite{head2014rafts, muriel2018rol}. 
        They are a major membrane protein site, and are central \cskl\ trafficking targets, interfacing with the actin \cskl\ \cite{sezgin2017mys}. 
MTB proteins, mainly tau, associate with LRs, and impairing this association yields axonal retraction \cite{whitehead2012rafts, head2014rafts}.

Palmitoylation targets proteins to LRs and is required for stable membrane anchoring of many proteins, including synaptic density proteins, glutamate receptors and their recycling proteins, steroid receptors, GPCRs, and actin regulating proteins required for spine growth \cite{albanesi2020palmitoylation}. 
 Caveolae are LRs that contain caveolin1 and show as membrane invaginations. They are essential for membrane anchoring and trafficking of tyrosine kinase receptors such as TrkA (NGF) and the insulin receptor \cite{pike2005gro}. 
 LRs and \cholest\ are essential for the activation of nicotinic receptors \cite{baenziger2017role}. 
 
LRs are essential for \plast. 
Synaptogenesis requires \astc -derived \cholest\ in ApoE \cite{mauch2001cholest}.  Depletion of \cholest\ or sphingolipids leads to gradual loss of \hippo\ dendritic spines and synapses and to surface AMPAR instability \cite{hering2003rafts}. 
 \Cholest\ depletion blocks winner enhancement in \hippo\ CA1 \cite{frank2008cholest}, an area that shows specific vulnerability in AD (see below).

LRs are present in intracellular membranes as well, including in mitochondria-associated endoplasmic reticulum membranes (MAMs), which are essential for \mitoch\ function \cite{area2012upr}, and in the trans-Golgi network and recycling endosomes \cite{waugh2013raf}.

 \msubsection{Competition resolution (\cres): winners, losers}
\cgen\ induces indiscriminate de\stab, candidate generation, and cellular calcium influx in the vicinity of activity sites. 
\Cholest\ induces a switch from \cgen\ to \cres, in two ways. It allows winner enhancement through LRs, and it stimulates the production of Abeta, which is the main loser removal agent.

\cres\ has four main components. First, calcium influx is reduced, in both winners and losers. 
Calcium must be restrained because it can be toxic.  Most relevant here, calcium activates calmodulin, whose continued binding to PSD95 rapidly (within 15 minutes) decreases its palmitoylation and synaptic location and increases AMPAR endocytosis, rendering synapses unusable \cite{chowdhury2019ca2}. This time frame accords with the duration of short-term memory. 
 Calcium is restrained by switching GluN2B to GluN2A \cite{sanz2013diversity}, replacing CP-AMPARs by CI-AMPARs by adding GluA2 \cite{plant2006tra}, and endocytosis of GluA1 via PP2B \cite{miller2014tau}. 
  
Second, the \cskl\ is linked to winners to support trafficking, with tau being a major player (see below). Third, winning synapses are stabilized, including the formation of \nanod s for efficient neuro\transm\ \cite{liu2021reg}. 
 Finally, the membrane is returned to a stable condition by removing losing candidates (see Abeta below).

 \smallest{Winner determination}
\catwo\ is a DEP agent, with high amounts (or rates) activating kinases (mainly CaMKII) that promote winners, and small amounts activating phosphatases (mainly PP2B) to eliminate losers \cite{shouval2010plast}.
 It is not possible to restrain calcium before some persistent enhancement to winners, because strong, focused, persistent calcium is what determines winner status. Once LRs are incorporated into the plasma membrane, \cres\ can enhance winning candidates and remove losing ones to establish a new steady state. 

There are two main agents involved in winner enhancement, CaMKII and PKC. Both start their activity during \cgen\ (because winners obviously need to be generated first) and continue during \cres. 

CaMKII is activated by high calcium \cite{nicoll2023syn} during \cgen. 
 It binds GluN2B, translocates with it to synapses, and promotes GluA1 channel conductance via \phos\ at S831 \cite{nicoll2023syn}. 
It induces spine growth via Rac1 \cite{saneyoshi2019rec}. 
 When \cres\ starts, CaMKII undergoes auto\phos\ at T286, which allows it to continue activation without calcium. This way, it can support \cres\ to \consol\ and possibly even synapse maintenance at the steady state. 
It promotes synapse \stab\ and \nanod s via segregation of AMPARs and NLG from GluN2B \cite{liu2021reg}, and the switch from GluN2B to GluN2A \cite{sanz2013diversity}. 
   
PKC is activated by sustained calcium from \icell\ stores.
  During \cgen, it promotes GluA1 synaptic delivery via S818 \phos\ \cite{boehm2006synaptic}, 
 structural growth via Rac1 \cite{pilpel2004activation}, 
  production of sAPPa over abeta \cite{buxbaum1990APPalpha}, 
 and GluA2 endocytosis \cite{chung2000pho}. 
 During \cres, it promotes the switch from GluN2B to GluN2A \cite{sanz2013diversity}, and prevents winner damage due to Abeta (endocytosed APP undergoes beta cleavage, unless this is opposed by PKC). 
 At \plast\ termination, its constitutively active form PKMz supports long-term memory \cite{sacktor2018doe}. 
 
  Candidates that become large enough allow the invasion of MTBs and tau \cite{egawa2016rafts}, an important factor in winner determination.

 \smallest{Tau}
Tau binds \cskl\ MTBs and was initially thought to support MTB stability. However, its detachment from MTBs does not destabilize them or impair axonal transport \cite{bakota2023kiss}. 
 It is now viewed as a cross-linker of the MTB and actin \cskl s that promotes dynamic changes including axonal elongation, synapse formation, and \stab\ of growing neurites \cite{cabrales2017mul}. 
  
During \cgen, tau is inactivated via \phos\ (see above) to oppose its \stab\ effect. 
 During \cres, it is activated in winners to promote their structural \stab\ and \cskl\ linking, and inactivated in losers to allow their removal. Activation in winners is done by phosphatases, mainly PP2A \cite{gong2008hyp}. 
 Methylated (stronger) PP2A is enriched in LRs and decreases p-tau \cite{sontag2013PP2A}.
 Inactivation in losers is mainly done by GSK3b and Cdk5, assisted by p75 \cite{shen2019neurotrophin}.

     As noted above, active tau requires LRs to achieve its effects \cite{whitehead2012rafts, head2014rafts}. 
  Reducing \cholest\ (which induces a loser-like, non-LR state) yields hyper \phos\ of tau at the GSK3 and PKA sites \cite{bai2021dhc}. 
  Excess tau is secreted via direct translocation through the plasma membrane, and this action is supported by \cholest\ and SPM \cite{merezhko2020cell}. Thus, \ecell\ tau is a normal by-product of \plast-inducing activity.

 \smallest{Abeta}
Abeta is a major \cres\ agent whose main role is to remove losers. The switch from alpha to beta cleavage of APP, and thus initiation of Abeta production, occur when LRs have been formed. 
          \Cholest\ clearly stimulates Abeta production and is needed for it, its depletion inhibiting BACE1 and gamma secretase additively \cite{simons1998cholest, grimm2008ind}. 
   In turn, Abeta induces negative feedback on \cholest\ synthesis \cite{grimm2005cholest}, supporting the \ptname\ view that \cholest\ synthesis occurs in \cgen\ but not \cres. 
    
BACE1 operates in \icell\ compartments (endosomes, Golgi), so Abeta can be produced without the formation of plasma membrane LRs \cite{cho2020pre} (as we will see below, this has important implications in AD). 
Plasma membrane LRs are a feature of winners, while the main role of Abeta is to remove losers, which are spatially segregated from winners.

 Abeta production is also promoted by the \cres\ agents p75 and JNK \cite{costantini2005TRF, guglielmotto2011JNK}. 
             Abeta promotes \cres\ by enhancing GSK3beta \cite{shen2019neurotrophin, jimenez2011APPalpha}, 
    PP2B \cite{hsieh2006ampar, miller2014tau}, 
    proNGF \cite{bruno2009amyloid}, 
 p75 \cite{shen2019neurotrophin}, 
  PTEN \cite{knafo2017pten}, 
 JNK \cite{guglielmotto2011JNK}, 
   and RhoA (which opposes Rac1) \cite{petratos2008bet}. 
 Abeta's loser removal functionality is obtained via three main mechanisms, RhoA, p-tau (via GSK3b, JNK) and p75, which is a high-affinity receptor for proBDNF and proNGF and a low-affinity receptor for NGF. p75 stops neurite growth and induces apoptosis \cite{park2013BDNF}, 
 and, with Abeta, activates sphingomyelinase to remove SPM from the plasma membrane \cite{grimm2005cholest}, thereby cleaning losing candidates from LR components and restoring membrane integrity. 
 
Abeta opposes \cgen\ through these agents (e.g., GSK3beta opposes Akt and thus insulin signaling, PP2B opposes BDNF, PTEN opposes mTOR) and other mechanisms (e.g., it opposes NGF \cite{jimenez2011APPalpha} and nAChRs, mainly the calcium-permeable alpha7 \cite{dougherty2003abeta}).

 The ECM, which promotes the stable post-\plast\ state, contains several agents (e.g., HSPG) that are contained in LRs and bind and uptake Abeta \cite{zhang2014towards}. 
    
Abeta is a DEP agent, possibly to prevent it from impairing winners. Abeta oligomers readily insert in membranes to form cation-selective pores \cite{dougherty2003abeta, diScala2014interaction}. 
      In low (picomolar) concentrations, the resulting calcium influx promotes the function of alpha7nAChRs \cite{dougherty2003abeta}, explaining the positive effect that low Abeta can have on LTP and growth \cite{puzzo2008pic}. 
 In higher (nanomolar) concentrations, Abeta inhibits nAChRs and decreases \plast\ \cite{dougherty2003abeta, puzzo2008pic}. 

In summary, Abeta terminates \cgen, promotes \cres, especially loser removal, and terminates \cholest\ synthesis. 
The common view of Abeta as a negative agent stems from the fact that loser removal involves reduced spine density and LTD. 
However, loser removal is essential for cellular health, and Abeta has a crucial healthy role in \plast. Indeed, Abeta disruption prevents the consolidation and stabilization of memory \cite{finnie2020amyloid}.

 \smallest{\Consol} 
This stage takes place after activity termination, and involves protein synthesis and trafficking to the winning candidates, completion of loser removal, cytoskeleton \stab, ECM and myelin production, and other actions needed to establish a stable steady state. \Consol\ is one of the major processes taking place during sleep \cite{rasch2013sleep}.

 \smallest{Summary} 
Brain \plast\ starts with de\stab\ of the stable state, coupled with indiscriminate increase of neural excitability and generation of structural candidates competing for enhancement. Winners are determined by sustained calcium signaling. Incorporation of \cholest\ in plasma membrane \lrf s triggers a switch to re\stab, which involves winner enhancement and loser removal. The major agents in these processes are the AD-associated agents tau, APP, and \cholest\ (with other lipids). sAPPa promotes the initial candidate generation stage. \Cholest\ induces a switch to the production of Abeta, which promotes loser removal. The formation of plasma membrane LRs allows activated tau to extend the \cskl\ and link it to the enhanced winning synapses. Both Abeta and tau \phos\ are crucial for normal, healthy \plast. 

 \msection{Lipid Rafts Theory of AD (\TAD)}

Provided with a theory of brain \plast\ that explains the roles of the main AD agents, it is now possible to understand what goes wrong in AD. This section presents the LRs theory of AD (\TAD) and shows how it accounts for AD symptoms and pathology. Risk factors, protective factors, and strong evidence supporting the theory are given in subsequent sections. Unless explicitly noted, the discussion is about sporadic AD, not FAD. 

 \smallest{Mechanisms underlying symptoms and pathology}
Our starting point for explaining AD are the following three clues. First, the strongest genetic risk factor in AD, the greatly increased risk in ApoE4 carriers, points to reduced transport, neural uptake, or neural trafficking of \cholest. 
Second, the characteristic symptom of AD, impaired anterograde memory with functioning short-term memory, points to a problem in the initiation of \cres\ or in winner enhancement. Third, the main AD pathologies, hyper p-tau and Abeta, point to chronic \cgen\ or excessive loser removal. Neuro\degen\ is a natural consequence of these pathologies and of chronic loser removal. 

The main role of \cholest\ in short-term \plast\ is in LRs. Thus, the first clue leads to the hypothesis of impaired LR formation. Impaired LR formation impairs the switch from STM to LTM and from \cgen\ to \cres, explaining memory symptoms (the second clue). LRs and \cres\ are essential for tau de\phos, explaining the tau part of the third clue (see below for the Abeta part). Neurons can produce \cholest\ in stress to support Abeta production, but they cannot supplant \astc-derived \cholest\ for plasma membrane LRs \cite{mauch2001cholest}. 
 Thus, we reach the following account: 
\itemb
\item AD symptoms and pathology are caused by impaired formation of plasma membrane lipid rafts, which is in turn caused by reduced neural uptake or trafficking of \astc-produced \cholest. 
\iteme

 \smallest{Chronicity}
The formation of PM LRs triggers the switch from \cgen\ to \cres. When it is impaired, we can expect both \cgen\ and \cres\ to be active at the same time, but with smaller agent amounts than those involved in healthy \plast, and with a prolonged duration. This is precisely what is called a {\me chronic} state (Figure \ref{fig4}). 
In early disease stages, chronicity should yield an increase in both \cgen\ and \cres\ agents. With disease progression, brain resources get depleted, so \cres\ should increase and \cgen\ decrease. This is exactly what the evidence shows (see below). 

Many \plast\ agents obey DEP, and are present in smaller amounts in chronicity. As a result, their effects are tilted towards \cres. Specifically, chronic calcium induces loser removal and neuro\degen. Thus, the continued activation of \cgen, which could have compensated for \cholest\ deficiency by promoting its synthesis, actually exacerbates the situation. 
 Moreover, because biological systems extensively rely on negative feedback, 
     chronicity desensitizes receptors and pathways, impairing many cellular signaling pathways.
 
 \smallest{\Cholest, Abeta}
A straightforward explanation for reduced PM LRs is decreased \cholest\ synthesis . Indeed, there is data indicating that ApoE4 is associated with reduced synthesis capacity \cite{lazar2022lip}, and evidence for reduced \cholest\ in AD (see below). 
 However, chronic synthesis is at least as likely, since \cholest\ synthesis continues as long as it does not get incorporated properly in the plasma membrane. Thus, AD can involve both \cholest\ deficiency (in PM LRs) and excess. Chronic synthesis happens in both \astc s and neurons, which increase \cholest\ production under stress. Neural production may be able to counter membrane damage incurred by injury \cite{mahley2012injury}, but it generally does not compensate for \cholest\ deficiency during \plast\ \cite{mauch2001cholest, ferris2017SREBP2}. 
     
   \Cholest\ stimulates Abeta production, yielding two types of neurons associated with chronic \cholest\ in AD. In the first type, chronic neural \cholest\ production eventually manages to form PM LRs. Such neurons form synapses and do not show tau pathology, but do show excessive Abeta production, causing Abeta secretion and aggregation. This explains why Abeta is common in aging, and why Abeta plaques are not strongly correlated with AD symptoms and tau pathology. In the second type, chronic \cholest\ production does not compensate for reduced PM LRs, and these neurons show \plast\ impairment and both tau and Abeta pathology. Since Abeta binds \cholest\ \cite{diScala2014interaction}, 
 in both neural types Abeta plaques are expected to contain \cholest, as is indeed the case \cite{stewart2017amyloid}.

 \smallest{Tau}
LRs are crucial for \cskl-PM linking, neurite growth, and neurite stabilization, and tau has a major role in these functions \cite{whitehead2012rafts, head2014rafts}. 
  LRs promote the switch to \cres\ and allow tau de\phos\ in winners. LR impairment promotes the cytosolic localization of tau and p-tau, and tau-membrane disconnect may facilitate the helical tau conformation found in NFTs \cite{georgieva2014tau}. 
  Propagating p-tau is hypothesized to spread AD pathology \cite{kaufman2018tau}, and tau is indeed secreted as a normal consequence of neural activity \cite{merezhko2020cell} (see above). 
 \ptname\ explains this tau secretion as a way of clearing excessive inactive tau generated during \cgen, but it is possible that it occurs in AD, which involves excessive inactive tau as well. Nonetheless, gradual spreading is not well-supported by the evidence \cite{jack2018longitudinal}. 
     
 \smallest{Core cause}
The mechanisms of AD can be explained by impaired PM LRs, and these can be explained by reduced uptake or trafficking of \cholest. Why does this happen? Although the link with ApoE4 is very strong, there are people carrying this isoform (even homozygotes) who do not get the disease. In addition, there are ApoE3 carriers with AD. 

The parsimonious answer to this question is that \cholest\ \metab, and specifically LRs, are normally reduced in aging \cite{martin2010cholest, egawa2016rafts}\footnote{Which incidentally explains why memory is reduced in normal aging.}, 
  and a variety of risk factors can push this reduction beyond the damage threshold.  Some viruses enter cells via LRs and impair lipid \metab\footnote{This includes {\sc covid-19}, which might explain its long-lasting cognitive symptoms.}. Insulin resistance and impaired vasculature impair all \plast\ stages and can specifically induce \cholest\ deficiency. Sleep problems, chronic \infl, and chronic stress promote a state similar to chronic \destab. Brain injury involves acute \destab\ that can become chronic. Polyunsaturated fatty acids (PUFAs) are also present in \lrf s and reduced in aging. These risk factors are expanded upon below. In addition, lipid \metab\ is supported by many genes, providing ample opportunity for small gene changes to predispose the person to less efficient \metab, especially in aging. Such changes can manifest as sporadic AD even in ApoE3 carriers. 

Gonadal steroids are major \plast\ agents and are reduced abruptly and relatively early in women (vs.\ a gradual loss in men), explaining why women have an increased AD risk at younger ages and a faster cognitive decline after diagnosis \cite{ferretti2018sex}.

 \smallest{Autosomal dominant AD}
The core cause of autosomal dominant AD (which approximately overlaps with familial AD) is different from that of sporadic AD, and involves decrease-of-function mutations in the PSEN1 gene, whose protein (presenilin 1) is a major component of gamma-secretase, or mutations in the gamma-secretase APP cleavage site \cite{bateman2011autosomal}. 
Such mutations affect both sAPPa and Abeta, which are central \plast\ agents. Hence, such mutations can seriously damage \plast, explaining the earlier age of onset of the disease. The reason that the brain manages to support \plast\ until middle age is probably the large functional redundancy between the various \cgen\ and \cres\ agents. 
  
 \section{Evidence}
In addition to the salient tau, Abeta, and ApoE4 evidence, many other lines of evidence provide support for \TAD, including from LRs, \cholest, PUFAs, \plast\ agents, synapse issues, the brain areas of earliest damage, genetics, and other diseases. 
These lines of evidence are presented in this section. Risk factors also provide support for the theory and are discussed in the next section. 

 \smallest{Lipid rafts}
LRs are reported to be reduced or abnormal in AD. In mixed ApoE3-ApoE4 carriers, membrane \cholest\ was decreased by 36\%, with abnormal LRs \cite{ledesma2003plasminAZ}. 
 Neuronal LRs are strongly altered in frontal cortex and ETRC from the earliest stages of the disease (before NFTs) \cite{fabelo2014alt}. The changes include lower \cholest, SPM, and PUFAs \cite{fabelo2014alt}. These neurons show increased sterol esters, which may indicate excessive \cholest\ synthesis with impaired incorporation into LRs. 
   Patient temporal cortex contains significantly lower LRs, with the remaining ones depleted of \cholest\ \cite{molander2005str}. Frontal and temporal cortex LRs contain significantly higher gangliosides GM1/2 \cite{molander2005str}. 
  Flotillin and gangliosides, strong \lrf\ markers, show marked abnormalities in patient brain, CSF and serum \cite{abdullah2019flo, mesa2022neu, chan2023rob}. In most cases, this also occurs in MCI patients, showing that it is an early event in the disease. 
   Flotillin accumulates in lysosomes in NFT-positive neurons, showing the link between LRs and tau pathology \cite{girardot2003acc}. 
          A recent review summarized strong evidence that Abeta plaques contain \cholest\ and gangliosides \cite{stewart2017amyloid}. 
 
 \smallest{\Cholest}
Dysregulated lipid metabolism, especially accumulation in glia, has been noted as a third major AD pathology right from the start, over 100 years ago \cite{foley2010AZ}. 
 \Cholest\ and lipid dysregulation is a strong characteristic of AD \cite{foley2010AZ}. 
  As predicted by \TAD, both lower and higher \cholest\ are reported. 
Lower \cholest\ is reported in temporal cortex \cite{mason1992evidence} and CSF \cite{yassine2016abc, kolsch2010alterations}, especially in severe patients \cite{phelix2011viv}. 
    Seladin1, an enzyme that mediates the conversion of desmosterol to \cholest, is reduced in AD, specifically in vulnerable regions \cite{bai2022rol}. 
 ETRC and \hippo\ show lower \cholest\ synthesis, which is significantly associated with AD pathology (however, \cholest\ levels themselves were normal, which might indicate chronicity) \cite{varma2021abn}. 
 
Most of the evidence points to chronically increased synthesis. 
      Neurons with NFT tau show higher unesterified \cholest\ \cite{distl2001tan}. 
 \Cholest\ and ceramides accumulate in AD brain in a vulnerable region (frontal cortex) but not the cerebellum \cite{cutler2004lipids}. 
 ETRC shows elevated \cholest\ esters, SPM, and ganglioside GM3 \cite{chan2012comparative}. 
 Patients show higher brain \cholest\ and decreased LXR (\cholest\ extrusion) \cite{xiong2008cholest}. 
 Increased ABCA1 (sterol extrusion) expression highly correlates with dementia severity in AD \hippo\ \cite{akram2010inc}. 
 \Cholest\ is increased in patient \astc\ \mitoch\ \cite{arenas2020sta}. 
 A sensitive imaging method shows significantly higher (34\%) \cholest\ in cortical layers 3 and 4 \cite{lazar2013cholest}. 
 AD pyramidal neurons in the \hippo\ show no nucleus translocation (activation) of SREBP2, indicating sterol excess. 40-50\% of NFT neurons express SREBP2, non-nuclear \cite{wang2019SREBP2}. 
      Total \cholest\ is increased in a CHO cell line with the FAD-causing presenilin1 mutation \cite{cho2019ele}. 
 CSF levels of \cholest\ correlate with p-tau \cite{popp2013cer} and secreted APP (both alpha and beta) \cite{popp2012cho} in patients but not controls. 
   PCSK9, which downregulates LDLR and thus ApoE uptake, was reported to be increased in AD, indicating excess. However, no changes have also been reported \cite{adorni2019pro}. 
    
Fibroblasts of sporadic AD and FAD patients show markedly higher free \cholest\ and \cholest\ esters \cite{montesinos2020alz}. 
 \Mitoch-associated endoplasmic reticulum membranes, which contain LRs, show a higher number of contacts and upregulated function (\cholest\ esterification, phospholipid synthesis) in sporadic and FAD fibroblasts and in mouse models \cite{area2012upr}. 
  Patient skin fibroblasts show increased \cholest\ esterification enzyme and decreased SREBP2, ABCA1 mRNA \cite{pani2009alt}. 
 In addition to showing \cholest\ excess, these results are important because they show that the problem can be present in cells other than glia and neurons, shedding light on the core cause and allowing practical biomarkers. 

  In many cases, the core of the problem may be lipid trafficking rather than uptake. ApoE4 (and to a lesser degree, ApoE3) tends to aggregate in and damage endosomes \cite{vanAcker2019endo}, 
 which are crucial for \cholest\ homeostasis. 
Indeed, endosomes are dysregulated in AD, and endosome-related genes such as Bin1, TREM2, PICALM, and SORLA have been highlighted in genome-wide association studies (GWAS) \cite{vanAcker2019endo}. 

 \smallest{PUFAs}
DHA, the main omega 3 PUFA, is reduced in AD serum and post-mortem brain \cite{grimm2017app}. PUFAs are the main target of lipid peroxidation by reactive oxygen species produced during oxidative stress, and AD brains show higher levels of PUFA oxidation products and oxidative stress (which can stem from chronicity, e.g., of calcium) \cite{grimm2017app}. 
Although PUFAs are mainly located outside LRs, some are located inside LRs, and PUFA levels are reduced in LRs in frontal cortex and ETRC in AD \cite{fabelo2014alt}.

 \smallest{\Plast\ agents}
Virtually all important \plast\ agents are dysregulated in AD, including \catwo\ \cite{liu2019his}, ERK, p38, JNK, Akt, PKA, PKC, GSK3beta, PP2B, PP1, PTEN, Cdk5 \cite{chung2009aberrant}, BDNF, TrkA (the NGF receptor), tPA, MMP9 \cite{lulita2016ngf}, and PAI-1 (tPA inhibitor) \cite{hebert2016sto}. 
   In this list, tau \phos\ agents (ERK, p38, JNK, GSK3, PP2B, Cdk5, but not PKA) are chronically increased \cite{chung2009aberrant}, while PP2A, the main tau de\phos\ agent, is reduced in AD \cite{gong2008hyp}. 
  
Notably, PKC \cite{chung2009aberrant} and GluA2 \cite{carter2004dif}, the hallmarks of winner enhancement and stabilized winners, are reduced in AD. 
 In the \hippo, the greatest decrease in GluA2 occurs in the most vulnerable areas (subiculum, CA1) \cite{carter2004dif}. 
 Note that PKC is activated by moderate calcium, so chronic calcium (e.g., induced by membrane Abeta cation channels) should have activated it, which would explain reduced GluA2 (PKC promotes its endocytosis, see above). The fact that PKC is reduced and not increased indicates a fundamental non-Abeta problem in winner enhancement processes.

 \smallest{Synapses}
Synaptic damage is a well-known early event in AD \cite{crews2010molecular}. This shows that synapses are not formed correctly or are degraded early. Both accounts support \TAD. 
 
 \smallest{Brain areas}
The earliest areas that show tau pathology in AD are the ETRC, \hippo, and LC \cite{kaufman2018tau}. 
    Among \hippo\ fields, the subiculum and CA1 are much more vulnerable than the DG and CA3 \cite{small2011pat}. 
 Among thalamic nuclei, the nucleus reuniens, connecting the ETRC, subiculum, CA1, and mPFC, is one of the most heavily affected nuclei \cite{braak1991alz}. 
 ETRC and \hippo\ represent scenes and events, which due to their richness present continuous novelty that requires \plast . The LC, the main norepinephrine center in the brain, is also active in all novel situations. The fact that the areas that show the largest \plast\ requirements are those most vulnerable in AD supports \ptname. Moreover, 
 CA1 and subiculum support familiar scenes, while DG and CA3 support the encoding of new ones \cite{maass2014lam}. Most of our life experiences involve familiar scenes with some element of novelty, explaining why CA1 and subiculum are relatively more vulnerable. ETRC supports both familiar and novel scenes, explaining its central role in early AD. 
  
Early auditory \cite{thomas2014loc}, olfactory \cite{brady2004fun} and retinal \cite{munderloh2009reg} sensory cells contain LRs, 
   explaining the early loss of hearing and smell and retinal abnormalities in AD \cite{murphy2019olf}. 
 Of note, statin use is associated with sudden hearing loss in an all-Taiwan insurance database \cite{chung2015pop}, 
 statins are among the drugs that can impair olfaction \cite{lotsch2015olf}, 
 and 
\cholest\ depletion causes hearing loss in cats and severe cochlear hair cell loss in mice \cite{crumling2012hea}.

 \smallest{Genetics}
\Cholest\ and lipid pathway genes are repeatedly identified as leading clusters in AD GWAS \cite{dong2017integrated}. So are endosome-related genes, as mentioned above.

 \smallest{Other diseases}
Niemann-Pick disease type C (NPC), which involves dementia and neurodegeneration, is caused by mutations in the NPC1 or NPC2 genes, which support transport of \cholest\ from late endosomes and lysosomes \cite{love1995NPC}. NPC brains show tau-rich NFTs with paired helical filaments identical to those found in AD \cite{love1995NPC}.
 
Down syndrome patients usually carry three copies of the APP gene, and manifest AD-like pathology \cite{castello2013rational}.

 \msection{Risk \& Protective Factors}
 Here I discuss various factors affecting AD. Many of these factors are associated with aging, increasing the aging-related risk. 

 \smallest{Viral infections} 
Increased human herpesvirus 6A and 7 were reported in AD \cite{readhead2018herpes}, and herpesvirus infection was found to induce a threefold increase in dementia risk \cite{tzeng2018herpes}. 
       Moreover, in a highly controlled all-Wales setting, herpes zoster vaccination was shown to decrease dementia (all cause except vascular) risk by 20\% \cite{eyting2023cau}. 
 
In \TAD, viral infection is a direct AD risk factor because many viruses (including herpesvirus and {\sc covid}-19) highjack lipid metabolism to persistently weaken lipid homeostasis, especially \lrf s \cite{itzhaki2006herpes}.
       Incidentally, this may explain the long-lasting cognitive symptoms of {\sc covid}-19.

 \smallest{Traumatic brain injury (TBI)}
TBI can induce a physical disconnect of the ECM, which triggers LR endocytosis \cite{muriel2018rol}. 
 Moreover, it involves strong release and prolonged presence of \cgen\ agents in the brain \cite{li2020update}, which can induce chronic loser removal due to DEP. 
Indeed, TBI is an AD risk factor and is associated with synaptic dysfunction 
\cite{li2020update}.

 \smallest{Diabetes, obesity}
Type 2 diabetes and obesity are AD risk factors \cite{AZ2023}, and the notion that AD is `type 3 diabetes' is gaining support \cite{deLaMonte201820}. In our view, this notion is not justified. \TAD\ explains this risk factor via the fact that the synthesis of lipids, including \cholest\ and PUFAs, needs PI3K/Akt signaling, which is impaired by insulin resistance (IR). Thus, IR can impair LRs and promote chronic \cholest. 
 
Although IR can exacerbate the situation of people who are at AD risk, brain IR is just one of many risk factors that can yield LR damage, which is the core mechanism in AD. 
Note that \cholest\ homeostasis is more fundamental in AD than IR, since if IR were the core cause it should have affected ApoE2 homozygotes, who are virtually protected. We conclude that IR is definitely a serious AD risk factor and it may even suffice for causing dementia, but the dementia classified as AD requires additional factors.

 It should be noted that the association between AD and obesity is not that simple, 
 since low BMI is also associated with increased risk 
\cite{qizilbash2015bmi}. 
 This supports \TAD, since low BMI may indicate a tissue-wide problem in lipid synthesis and/or uptake, which can impair LRs. 
    
 \smallest{Vascular impairment} 
Vascular dementia is the 2nd most common dementia after AD, accounting for 5-10\% of the cases, and cerebrovascular disease is commonly shown in AD \cite{AZ2023}. Cardiovascular disease is also associated with increased AD risk \cite{AZ2023}. 
 Vascular impairment can increase AD risk via reduced raw materials needed for lipid synthesis. However, it can also be a consequence of the AD core mechanism, since impaired \cholest\ homeostasis yields vascular damage.

Statins suppress \cholest\ synthesis and are regularly taken by people at perceived risk of cardiovascular disease. They do not cross the blood-brain barrier so should not directly affect AD, but they might reduce risk via the other risk factors. They can also be detrimental in AD in case of barrier damage.

 \smallest{Immunity} 
Neuro\infl\ is common in AD, associated with activation of microglia \cite{liu2019his, AZ2023}
 and of their TREM2 receptors \cite{ulrich2017elucidating}. 
 TREM2 is activated by lipids, with mutations that increase AD risk and mutations that decrease TREM2-ApoE binding \cite{ulrich2017elucidating}. Microglia might be activated by \cholest\ excess and have an initial protective role in AD. However, chronic neuroinflammation promotes neurodegeneration.

 \smallest{Stress}
Early life and life-time stress are associated with earlier AD onset and faster disease progression \cite{hoeijmakers2018preclinical}. 
 This agrees with the fact that chronic stress is associated with neurodegeneration in memory-related areas \cite{stein2019spines}. 
  The mechanisms of damage of chronic stress may be related to LRs (e.g., cortisol induces endocannabinoid synthesis from PUFAs, which can impair LRs) or be independent of LRs (e.g., calcium excitotoxicity). As a neurosteroid, cortisol requires \cholest\ for its synthesis, so chronic cortisol synthesis may promote brain \cholest\ depletion. 

Oxidative stress and its lipid peroxidation effect are higher in AD, see PUFAs above.

 \smallest{Smoking}
Smoke exposure and smoking are associated with increased AD risk \cite{AZ2023}. This can occur via several mechanisms. For example, 
nicotinic agonists increase p-tau \cite{hellstrom2000inc}.

 \smallest{Education, bilingualism} 
Formal education   is a clear AD protective factor \cite{AZ2023}, and so is (less clearly) bilingualism \cite{paulavicius2020bilingualism}.
   \TAD\ explains these data as follows. Both involve memory-related learning, which creates synapses. A denser synaptic space might reduce the number of candidates, which is a function of novelty. It should facilitate the formation of new synapses, because synapse candidates are closer to each other and are thus easier to connect. Thus, the growth needs of the educated brain are smaller, so smaller amounts of \cholest\ are needed for \plast. Indeed, bilinguals have a larger gray matter volume \cite{heim2019bilingualism} and increased connectivity  \cite{perani2017impact}.     
  
 \smallest{Exercise} 
Most (not all) data indicate that exercise weakly reduces AD risk \cite{meng2020exercise}.          Exercise has positive effects on many of the risk factors discussed here, including brain vasculature, blood flow, brain clearance from chronic agents, obesity, sleep, and cardiovascular disease. 

 \msection{Discussion}
This paper presented the first complete theory of AD, complete in the sense that it mechanistically explains all of the major AD phenomena, including etiology, symptoms, pathology, and risk factors. To do this, the paper presented a new theory of brain \plast\ (\ptname ) that explains memory formation and the roles in health and disease of sAPPa, Abeta, tau, growth factors, kinases, phosphatases, \cholest, and other agents. \ptname\ enables the development of a coherent theory of AD. 

 \smallest{Related theories}
        \Cholest\ has been repeatedly recognized as an important factor in AD \cite{puglielli2003alzheimer, foley2010AZ}. 
 However, this was based on its promotion of Abeta, its link with CVD, and evidence of lipid dysregulation in AD. 
   I found only one paper that points to \cholest\ as the central agent in AD \cite{castello2013rational}, but it only focuses on chronic \cholest\ as the driving force behind neuro\degen, without addressing anterograde memory symptoms and \plast\ agents and processes. 
 All of this work views \cholest\ negatively (opposite to \ptname ) and does not explain the disease mechanisms and pathological process.

Likewise, tau has long been known to be the main pathological factor correlated with symptoms, but no coherent story for why it is phosphorylated in health and disease has been presented. The \ptname\ account in which  \cgen\ requires destabilization and \cres\ removes losing candidates naturally explains p-tau and its chronicity-induced accumulation.  
Many hypotheses have been raised in AD research, including cholinergic, Abeta, tau propagation, \mitoch, calcium, neurovascular, \infl, microbes, metal ions, and lymphatic hypotheses \cite{liu2019his}, and type 3 diabetes \cite{deLaMonte201820}. As discussed above, these are factors that affect AD risk but are not fundamental to the disease.

 \smallest{\ptname }
 Brain \plast\ is one of the most heavily researched areas of neuroscience. 
Competition between spines and elimination of losing ones is an established idea in \plast\ research, with strong evidence \cite{stein2019spines}. 
  \ptname\ is the first theory of \plast\ that presents a coherent mechanistic story as to how initial \potn\ switches to \consol, and the first that seamlessly integrates structural \plast\ with neuro\transm\ \plast\ and explains the roles of \cholest, APP, tau, and other agents. 

 Strong high-level support for \ptname\ comes from its structural similarity to other biological theories. Notably, candidate generation followed by competition is the cornerstone of the theory of evolution. In immunity, clonal selection of B and T cells can be viewed as candidate generation,  while memory T cells result from consolidation.  
The principle of DEP has some superficial resemblance to the notion of hormesis in toxicology \cite{mattson2008hor}. 
 While the opposite effects of CaMKII vs.\ PP2B and of TrkB/TrkA vs.\ p75 are clearly known, as far as I know this paper is the first in which DEP is phrased as a general biological principle. Note that DEP applies to neuromodulators as well (see dopamine and serotonin examples given above). It also applies to the immune system, with mechanisms that will be described elsewhere.

This paper focused on the main notions relevant to \plast\ and AD, to show the forest rather than a huge number of trees. Additional topics are left to future texts, including the difference between LDLR and LRP1, other LRPs, gangliosides and SPM, Src and Fyn kinases, prion protein, reelin, AICD, smoking, myelin, mGluRs, Cdc42, Arc/Arg3.1, adenosine, p38, cofilin and other non-tau \cskl\ proteins, specific adhesion molecules, alpha-synuclein, TGF-beta, NFkB, homeostatic \plast, the difference between different Abeta species, etc. 
         
A topic that is of particular importance is the timeline of winner enhancement. For example, CaMKII is autophosphorylated to keep it going without calcium. Does PKC (which is calcium-dependent) switch to PKMz (which shows constitutive activity) at the same time? It seems that it should, otherwise it would continue to promote sAPPa and oppose GluA2. 
In addition, is GluN2A eventually removed from winners, like GluA1? 
More research is needed on these topics. 

Additional research is also needed with respect to LR homeostasis. How precisely does \cholest\ get incorporated in PM LRs? One possibility is that it is transported by caveolin1, but the precise mechanisms are still not clear.

 \smallest{Sleep}
AD is associated with impaired sleep. Sleep is crucial for memory and \plast, and the most upregulated genes in sleep support lipid synthesis and transport and LRs \cite{rasch2013sleep}. There is some controversy around the synaptic homeostasis hypothesis, in which the goal of sleep (or at least of slow-wave sleep) is to globally downscale synapses. In \ptname, sleep supports the late part of \cres, which indeed involves massive loser removal and reduced calcium entry in winning synapses. However, sleep also strengthens winning synapses.

 \smallest{\Ngns\ (NGNS)}
Adult NGNS is crucial for memory \cite{goncalves2016adult}, and occurs in the hippocampus DG, which receives inputs from ETRC reelin neurons, which show specific vulnerability in AD. 
In \ptname, new neuron candidates are generated during \cgen. New cells require substantial resources to mature, so we expect adult NGNS to be impaired in AD, which is indeed the case \cite{moreno2019NGNS}. 
    Unlike synaptogenesis, which is a ubiquitous process, NGNS-related evidence is small. However, its impairment may well be important in AD. 

 \smallest{Dementia}
This paper focused on AD, but \ptname\ and \TAD\ also provide a reasonable account of other types of dementia, in particular vascular dementia. In addition, zinc, iron and copper are fundamental in cellular processes and are dysregulated in AD \cite{liu2019his}. Metals interact with ApoE, and the link to LRs and to other types of dementia should be further investigated. 

Frontotemporal dementia (FTD) has a major variant that shows tau pathology. Nonetheless, in my view, both the core cause and the disease mechanisms of FTD are different from those of AD. A detailed theory of FTD will be presented elsewhere. 

 \smallest{Theory predictions}
 \ptname\ and \TAD\ are supported by very strong evidence, including by patient data and by a large number of animal models developed for AD and brain \plast. This evidence pertains to the details underlying the theories. Our main theory prediction is that what causes the human disease diagnosed as AD is reduced \cholest\ in neural PM LRs during \plast. This has not been directly shown yet. One way of showing it would be via clinical trials with the participation of people diagnosed with AD or known to be at a high risk, which is obviously a very long and risky process. 
   
The presented theory points to several promising directions for treating AD. I hope that some of these will be taken, to help patients, families, and caretakers. 

 \section*{Acknowledgements}
I thank Nimrod Rappoport for many discussions on various biological matters, and for commenting on an early draft of this paper. I am grateful to Yair Safrai and Yoav Lorch for their continued support.

 \bibliographystyle{vancouver}

\newpage

\section*{List of Abbreviations}
Terms introduced in this paper are marked by a {\bf bold} font. 

24SHC: 24S-hydroxycholesterol (oxysterol).

27HC: 27-hydroxycholesterol.

a7nAChR: alpha7 subunit of the nicotinic ACh receptor. 

ABCA1: ATP-binding cassette transporter sub-family A 1. 
ABCG1: ATP-binding cassette transporter sub-family G member 1.

Abeta: Amyloid beta.

ACh: Acetylcholine.

AChE: Acetylcholinesterase (ACh breakdown).

AD: Alzheimer's Disease.

AICD: Amyloid precursor protein intracellular cytoplasmic/c-terminal domain.

AMPAR: Alpha-amino-3-hydroxy-5-methyl-4-isoxazolepropionic acid receptor.

ApoE (ApoE2, ApoE3, ApoE4): Apolipoprotein E (E2, E3, E4 isoforms).

APP: Amyloid precursor protein.

{\bf \ptname : theory of adaptive response \plast. }

BACE1: Beta-site amyloid precursor protein cleaving enzyme 1.

BDNF: Brain-derived neurotrophic factor.

BIN1: Myc box-dependent-interacting protein 1 (also Amphiphysin 2).

BMI: Body mass index.

CaMKII: \catwo/calmodulin-dependent protein kinase II.

Cdc42: Cell division control protein 42. 

Cdk5: Cyclin dependent kinase 5.

{\bf Cgen: \ptname\ candidate generation stage.}

CP: Calcium-permeable.

{\bf Cres: \ptname\ competition resolution stage.}

CVD: Cardiovascular Disease.

D1R, D2R: Dopamine 1 (2) receptor.

{\bf DEP: Double-edged \plast.}

DG: Dentate gyrus.

DHA: Docosahexanoic acid (omega 3 PUFA). 
 
ECM: Extracellular matrix.

ERK: Extracellular signal-regulated kinase.

ETRC: entorhinal cortex.

GABA: Gamma-Aminobutyric acid.

GluA1, GluA2: AMPAR subunits A1 and A2. 

GluN2A, GluN2B: NMDAR subunits 2A and 2B.

GM1: monosialotetrahexosylganglioside.

GPCR: G protein-coupled receptor.

GSK3: Glycogen synthase kinase 3.

GTPase: Guanosine triphosphate hydrolase enzyme.

GWAS: Genome-wide association study.

 HSPG: Heparan Sulfate ProteoGlycan.

IR: Insulin resistance.

JNK: c-Jun N-terminal kinase.

LC: locus coeruleus. 

LDL(R): Low-density lipoprotein (receptor).

LR: Lipid raft.

{\bf \TAD: Lipid rafts theory of AD.}

LRP-1: LDL receptor-related protein 1 (also alpha-2-macroglobulin receptor).

LTD: Long-term depotentiation.

LTP: Long-term Potentiation.

LXR: Liver X receptor.

mGluR: Metabotropic glutamate receptor.

MMP9: Matrix metallopeptidase 9.

MTB: Microtubules.

mTOR: Mammalian target of rapamycin.

nAChR: Nicotinic ACh receptor.

NFkB: Nuclear factor kappa-light-chain-enhancer of activated B cells.

NFT: Neurofibrillary tangles.

NGF: Neural growth factor.

NMDAR: N-Methyl-D-aspartic acid receptor.

p-tau: phosphorylated tau.

PAI-1: Plasminogen activator inhibitor-1.

PCSK9: Proprotein convertase subtilisin/kexin type 9.

PFC: Pre-frontal cortex.

PKA: Protein kinase A.

PKC: Protein kinase C.

PLC: Phospholipase C.

PM: plasma membrane. 

PP1: Protein phosphatase 1.

PP2A: Protein phosphatase 2.

PP2B: Protein phosphatase 3 (also calcineurin).

proBDNF, proNGF: Precursor forms of BDNF, NGF.

PTEN: Phosphatase and tensin homolog.

PUFA: Polyunsaturated fatty acid.

Rac1: Ras-related C3 botulinum toxin substrate 1.

RhoA: Ras homolog family member A.

sAPPa: soluble APP alpha (APP after alpha- and gamma-secretase cleavage). 

SNP: Single-nucleotide polymorphism.

SORLA: sorting protein-related receptor (also L11 and sortilin-related receptor with A-type repeats).

SPM: sphingomyelin. 

SREBP: Sterol regulatory element-binding protein.

TBI: Traumatic brain injury.

tPA: Tissue plasminogen activator.

TREM2: Triggering receptor expressed on myeloid cells 2.

TrkA, TrkB: Tropomyosin receptor kinase A, B.

\newcommand{\captionspace}{{\vspace{0.8cm}}}

\clearpage
\begin{figure*}
\begin{center}
\includegraphics[width=5in]{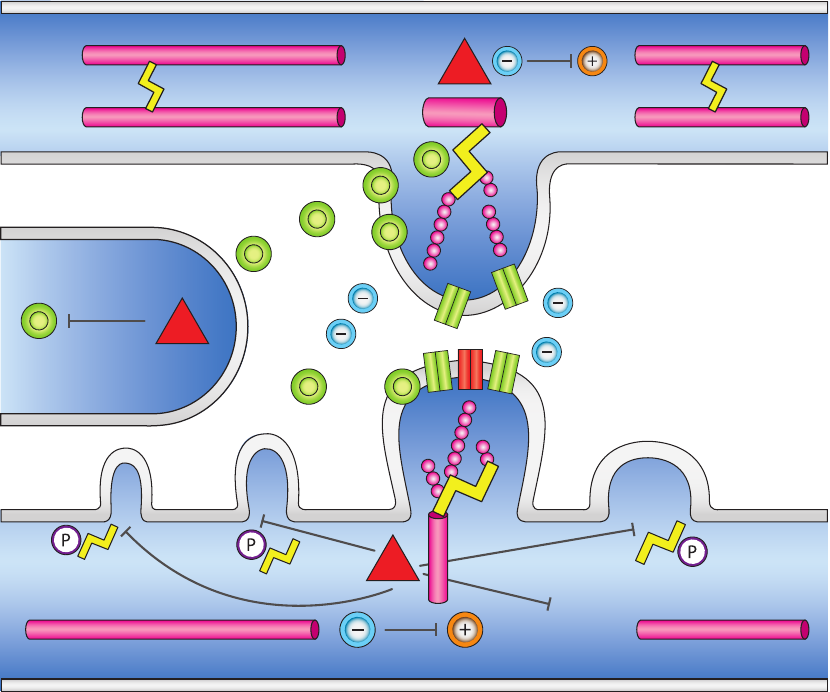} \captionspace

 \caption{{\bf Competition resolution (\cres).} \Astc-produced \cholest\ transported by ApoE (green circles) triggers the switch to \cres, which includes Abeta production (red triangle) and lipid raft formation. Winners are enhanced, active tau supports growth by linking microtubules and active actin, calcium-permeable glutamate receptors translocate to synapses, and Abeta suppresses losing candidates, \cholest, and \cgen\ agents. 
}
\label{fig2}
\end{center}
\end{figure*}

\clearpage
\begin{figure*}
\begin{center}
\includegraphics[width=5in]{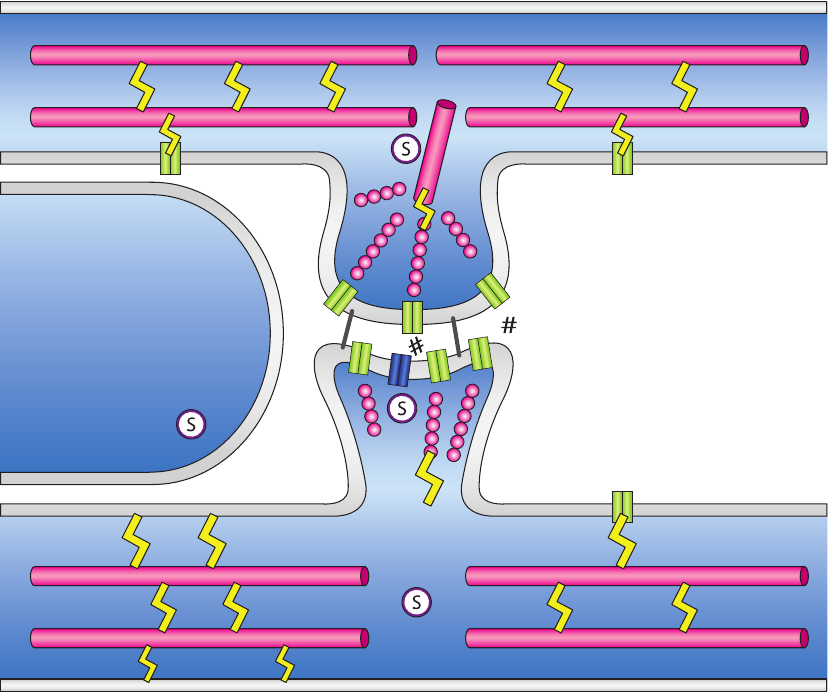} \captionspace

\caption{{\bf The stable state.} Presynaptic and postsynaptic winners and the \astc\ processes supporting the new synapse complete their growth. Stability agents (S circles) maintain the stable state. PP2A dephosphorylates tau to allow it to support structural elements. Lipid rafts cooperate with the cytoskeleton to support receptor anchoring. Synaptic glutamate receptors are switched to be calcium-impermeable (blue cylinders). Adhesion molecules (black lines) connect the two sides of the new synapse and form nanodomains, with ECM (\#) protection.
}
\label{fig3}
\end{center}
\end{figure*}

\clearpage
\begin{figure*}
\begin{center}
\includegraphics[width=5in]{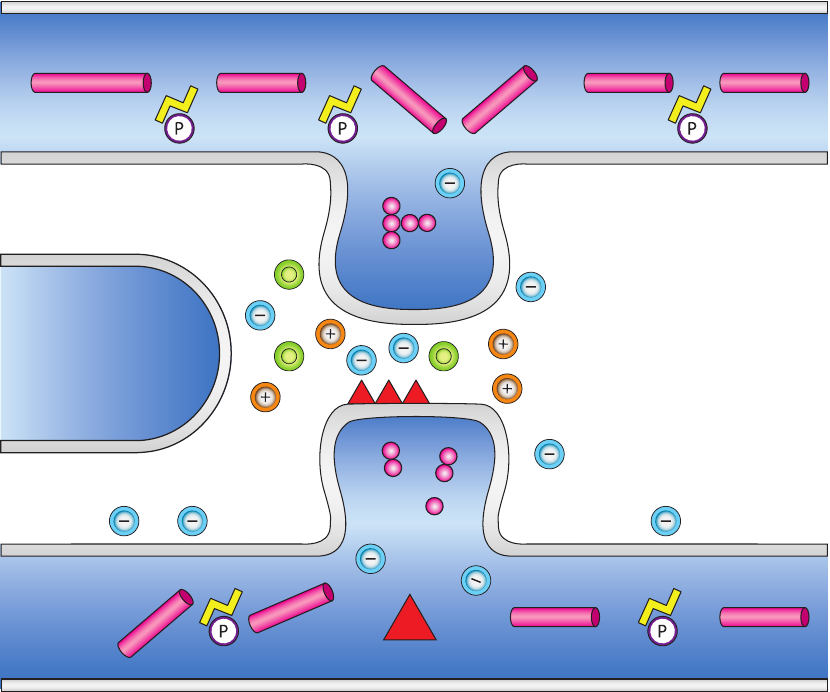} \captionspace

\caption{{\bf Chronic \plast\ and impaired \cres\ in AD.} In AD, \cholest\ is not incorporated sufficiently in plasma membrane \lrf s (e.g., due to reduced transport by ApoE4). As a result, synapse formation is not completed, manifesting as anterograde amnesia. LR deficiency impairs the switch from \cgen\ to \cres, and induces chronicity of their agents. Chronicity is biased towards \cres\ functionality due to the double-edged \plast\ (DEP) principle. This yields hyper p-tau, neurite retraction, cytoskeleton collapse, and eventually cell death, which manifest as retrograde amnesia. Chronic \cholest\ induces Abeta plaques, and Abeta forms toxic cation pores. (Intracellular \cholest, endosome issues are not shown.)
}
\label{fig4}
\end{center}
\end{figure*}
\clearpage

 \end{document}